\documentclass[cits]{PoS}

\title{Swift and Supergiant Fast X-ray Transients: a novel monitoring approach}

\ShortTitle{Monitoring SFXTs with {\it Swift}}

\author{ \speaker{P.\ Romano},$^a$ L.\ Sidoli,$^b$  V.\ Mangano,$^a$
J.A.\ Kennea,$^c$ G.\ Cusumano,$^a$ S.\ Vercellone,$^{ba}$ H.A.\ Krimm,$^{def}$ D.N.\
Burrows,$^c$ N.\ Gehrels,$^f$ \\
\llap{$^a$}INAF, Istituto di Astrofisica Spaziale e Fisica Cosmica, \\
         Via U.\ La Malfa 153, I-90146 Palermo, Italy\\
\llap{$^b$}INAF, Istituto di Astrofisica Spaziale e Fisica Cosmica, \\
         Via E.\ Bassini 15,   I-20133 Milano,  Italy\\
\llap{$^c$}Department of Astronomy and Astrophysics, Pennsylvania State  University, \\
         University Park, PA 16802, USA\\
\llap{$^d$}CRESST/Goddard Space Flight Center, Greenbelt, MD, USA\\
\llap{$^e$}Universities Space Research Association, Columbia, MD, USA\\
\llap{$^f$}NASA/Goddard Space Flight Center, Greenbelt, MD 20771, USA\\
E-mail: \email{romano@ifc.inaf.it}
}

\abstract{We describe our monitoring strategy which best exploits the
  sensitivity and flexibility of {\it Swift} to study the long-term
  behaviour of Supergiant Fast X-ray Transients (SFXTs). 
  We present observations of the recent outbursts from two objects of this class. 
  IGR~J16479--4514, underwent an outburst on 2008 March 19, reaching a
  peak  luminosity of about $6\times10^{37}$ erg s$^{-1}$ 
  (0.5--100 keV; at a distance of 4.9\,kpc). We obtained a  
  simultaneous broad-band spectrum (0.3--100 keV), the first for the
  SFXT class, which is fit with a heavily absorbed 
  (column density $5\times10^{22}$ cm$^{-2}$) hard power-law with a high
  energy cut-off at about 7\,keV. This spectrum shows properties 
  similar to the ones of accreting pulsars, although no X-ray
  pulsations were found. 
  IGR~J11215--5952, one of the only two  periodic SFXT known to date, was observed 
  with {\it Swift}  several times, first with an intense 23-day long 
  monitoring campaign around the 2007 February 9 outburst; then with a 
  26-day long monitoring around the unexpected July 24 outburst; 
  finally with a deep exposure during the 2008 June 16 outburst. 
  We present the whole dataset, which also includes observations 
  which allowed us to firmly establish the outburst period at
  $P\sim165$   days.
Thanks to our combined observations common characteristics to this class of objects are
emerging, i.e., outburst lengths well in excess of hours, often with a
multiple peaked structure, dynamic range $\sim3$ orders of magnitude,
and periodicities are starting to be found. 
}

\FullConference{7th INTEGRAL Workshop\\
		 September 8-11 2008\\
		 Copenhagen, Denmark}

\begin{document}

\section{Monitoring SFTXs with Swift}

Supergiant Fast X--ray Transients (SFXTs) are a new class of High Mass
X--ray Binaries (HMXBs)
discovered by INTEGRAL while monitoring the Galactic plane. They
display outbursts which are significantly shorter than typical
Be/X-ray binaries, peak luminosities in the order of  a few
10$^{36}$~erg~s$^{-1}$, and a quiescent level of  $\sim 10^{32}$~erg~s$^{-1}$. 
Since their spectral properties are reminiscent of those of accreting
pulsars, it is assumed that all the members of the new class are 
HMXBs hosting a neutron star, although the only 
three SFXTs with a measured pulse period are IGR~J11215--5952
($P_{\rm spin}\sim187$\,s, \cite{Swank2007}),  
AX~J1841.0$-$0536/IGR~J18410$-$0535 ($P_{\rm spin}\sim4.7$\,s, \cite{Bamba2001}), and 
IGR~J18483--0311 ($P_{\rm spin}\sim21$\,s, \cite{Sguera2007}).
The actual mechanisms responsible for the observed short outbursts
are still being debated, and the proposed explanations (see
\cite{Sidoli2008:cospar} for a review) 
involve either the structure of the wind from the supergiant companion 
\cite{zand2005,Walter2007,Negueruela2008}
or gated mechanisms (see \cite{Bozzo2008}). 

As we reported in \cite{Romano2007,Sidoli2007}, our {\it Swift} \cite{Gehrels2004}
monitoring of the outburst of 
the periodic SFXT IGR~J11215$-$5952 in February 2007 represents the
most complete and deep set of X--ray observations of an SFXT outburst. 
We discovered that the accretion phase during the bright outburst lasts longer than  
previously thought: days instead of hours, with only the brightest
phase lasting less than one day.

Given the success of these {\it Swift} observations, we extended the
investigation to a small although well-defined sample of SFXTs for
which a clear periodicity in the outbursts recurrence had not been 
found yet--as, on the contrary, was the case for
IGR~J11215--5952--with a monitoring campaign that would 
span the length of at least a few periods, i.e.\ in the order of 100--300\,days. 
The primary goal was to test whether our model \cite{Sidoli2007}
for the periodic SFXTs would also apply to some of
the other members of the class. 
Within this model, based on the presence of a disk-like wind component 
from the supergiant donor, inclined with respect to the orbital plane,
the outbursts are produced when the neutron star crosses the wind
component along the orbit. This model implies a periodicity or a
semi-periodicity in the outburst recurrence.

With its {\it unique fast-slewing and flexible observing scheduling},
which makes a monitoring effort cost-effective, 
its {\it broad-band energy coverage} that would allow us to model the 
observed spectra simultaneously in the 0.3--150\,keV energy range,
thus testing the prevailing models for accreting neutron stars,
and the {\it high sensitivity in the soft X-ray regime}, where some of the
SFXTs had never been observed, 
{\it Swift} was the most logical choice to monitor the light curves of our sample 
in order to search for the outburst recurrence predicted by our model. 

Our targets (IGR~J16479$-$4514,  XTE~J1739--302/IGR~J17391$-$3021, 
IGR~J17544$-$2619, and AX~J1841.0$-$0536/IGR~J18410$-$0535)
were selected considering sources which,
among several SFXT candidates,  are
confirmed SFXTs, i.e.\ they display both a ``short'' transient (and recurrent) 
X--ray activity and they have been optically identified with
supergiant companions (see references in \cite{Walter2007}).
In particular, XTE~J1739--302 and IGR~J17544$-$2619 are generally 
considered prototypical SFXTs. 
XTE~J1739$-$302 was the first transient which showed
an unusual X--ray behaviour \cite{Smith1998:17391-3021}, only recently optically associated
with a blue supergiant \cite{Negueruela2006}.
IGR~J16479$-$4514 has displayed  a more frequent X--ray outburst occurrence than other
SFXTs \cite{Walter2007} and offers an {\it a priori} better chance to be caught during an outburst.   
AX~J1841.0$-$0536, on the other hand, is an interesting source 
which may offer the opportunity  to determine the orbital parameters
from the pulsar timing on long time scales.

For these sources we requested 2--3 observations week$^{-1}$ object$^{-1}$, 
each 1\,ks long with {\it Swift}/XRT \cite{Burrows2005:XRT} in AUTO mode, to best exploit 
XRT automatic mode switching \cite{Hill04:xrtmodes} 
in response to changes in the sources' observed count rates. 
This observing pace would naturally fit in the regular observation scheduling 
of $\gamma$-ray bursts (GRBs), which are the the main observing targets for 
{\it Swift}. We also planned to propose for further target of opportunity (ToO)
observations whenever one of the sources showed interesting activity,
(such as indications of an imminent outburst)  or
underwent an outburst, thus obtaining a finer sampling of the light
curves and allowing us to study all phases of the evolution of an
outburst. 

With this setup, we aimed at fully characterizing the long-term behavior of 
SFXTs, to determine the properties of their quiescent state (where the accumulation of
large observing time is needed to allow a meaningful spectral analysis of this faintest emission), 
to monitor the onset of the outbursts and to measure the outburst
recurrence period(s) and duration.

\begin{figure}
\includegraphics[width=1.1\textwidth,height=0.97\textheight,angle=0]{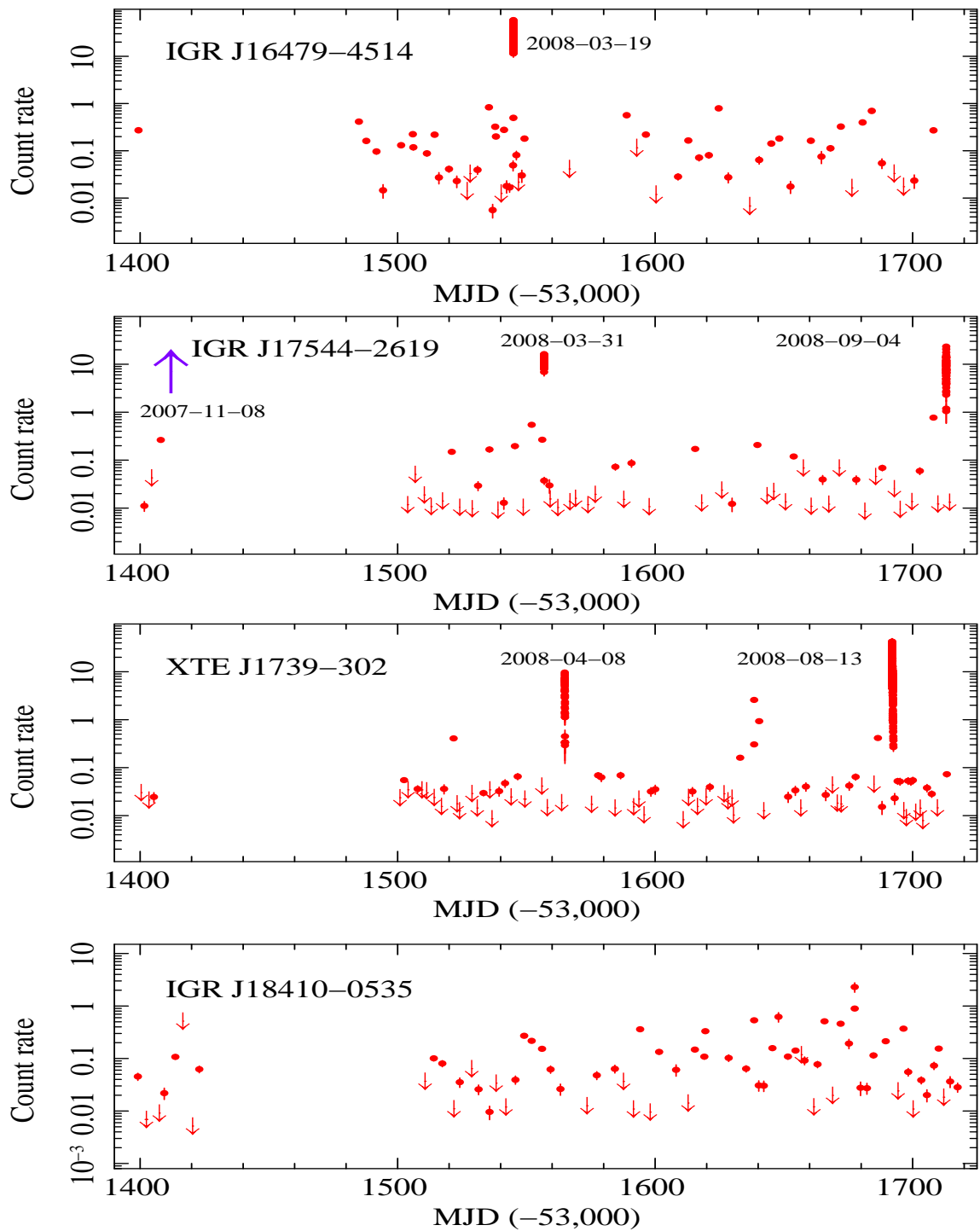}
\vspace{-2.5truecm}
\caption{{\it Swift}/XRT light curves of our sample of 4 SFXTs in the
  0.2--10\,keV energy range, between 2007 October 26 and 2008
  September 10. The light curves are background subtracted, and
  corrected for  pile-up 
  (when required), PSF losses, and vignetting. In each
  panel we report the dates of the observed outburst. Downward-pointing
  arrows are 3-$\sigma$ upper limits, while the upward pointing arrow in the
  light curve of IGR~J17544--2619 marks the 2007-11-08 outburst that 
  XRT could not observe because the source was Sun-constrained. The
  gap in the data points between November 2007 and January 2008 is due
to the sources being Sun-constrained.}
\label{fig1}
\end{figure}

Up to September 10 2008, we have collected a total of 306\,ks
distributed as shown in Table~\ref{tab1}. The XRT light
curves of the campaign are shown in Figure~\ref{fig1}. 
The long-term X--ray
emission outside the bright outbursts is described in full in
\cite{Sidoli2008:sfxts_paperI},
while the first observed outbursts of IGR~J16479$-$4514,
IGR~J17544$-$2619, and IGR~J17391$-$3021, 
are reported on in \cite{Romano2008:sfxts_paperII} and
\cite{Sidoli2008:sfxts_paperI},
respectively. Furthermore, in these Proceedings \cite{Sidoli2008:igr08} we report the 
preliminary results on two more outbursts of IGR~J17544$-$2619 and
XTE~J1739$-$302.
Here we shall highlight the results on the 2008 March 19 outburst of  
IGR~J16479$-$4514 and on a {\it Swift} GI observation on the 
2008 June 16 outburst of IGR~J11215--5952.

\begin{table}
\begin{tabular}{lrrrrll}
\hline
\hline
Name &Campaign Start &Number of    &{\it Swift}/XRT &Outburst &References \\
     &               &Observations & Exposure (ks)  & Dates   & \\
\hline
IGR~J16479--4514 &  2007-10-26	& 58&	67&	2008-03-19&    \cite{Romano2008:sfxts_paperII}\\
XTE~J1739--302 	 &  2007-10-27	& 83&  104&	2008-04-08&	\cite{Sidoli2008:sfxts_paperIII}\\
                 &   	        &   &     & 	2008-08-13&	\cite{Romano2008:atel1659,Sidoli2008:igr08}\\
IGR~J17544--2619 &  2007-10-28	& 57&	58&	2007-11-08&	\cite{Krimm2007:ATel1265}\\
  		 &    	        &   &     & 	2008-03-31&      \cite{Sidoli2008:sfxts_paperIII}\\
  		 &   	        &   &     & 	2008-09-04& 	\cite{Romano2008:atel1697,Sidoli2008:igr08}\\
AX~J1841.0--0536 &  2007-10-26	& 71&	77&	none      &     \cite{Cusumano2009} \\
\hline
\end{tabular}
\caption{Status of the {\it Swift} monitoring campaign as of September
  10, 2008.}
\label{tab1}
\end{table}

\section{Rise to the outburst in IGR~J16479$-$4514}

\begin{figure}[t]
		\vspace{-1.5truecm}
                \centerline{\includegraphics[width=9.5cm,height=13.5cm,angle=0]{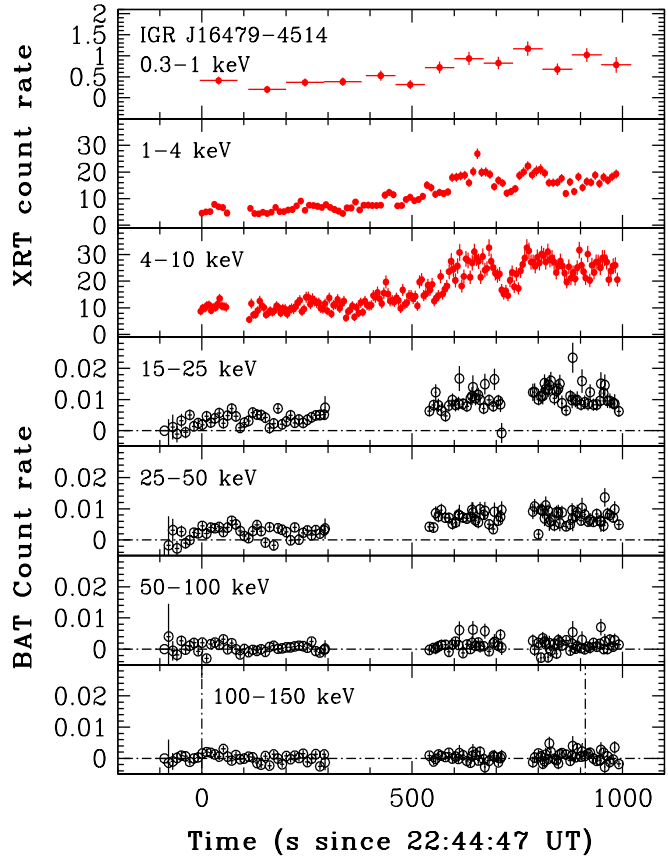}}	
		\vspace{-0.7truecm}
                 \caption{XRT (red filled circles) and BAT (black
                   empty circles) light curves of the 2008 March
                   19 outburst of IGR~J16479$-$4514
		in units of count s$^{-1}$ and count s$^{-1}$
                detector$^{-1}$, respectively. 
                The vertical dot-dashed lines in the bottom panel mark the two BAT triggers. 
		The XRT points up to $\sim 60$\,s after the first BAT trigger were collected
		 as a pointed observation part of our monitoring program. 
		The gaps in the BAT data are caused by BAT event mode time
		intervals being limited to less than 600\,s to reduce telemetry. 
		}
                \label{sfxts2:fig:lcv_allbands}
       \end{figure}

\begin{figure}[t]
                \centerline{\includegraphics[width=6cm,height=8.5cm,angle=270]{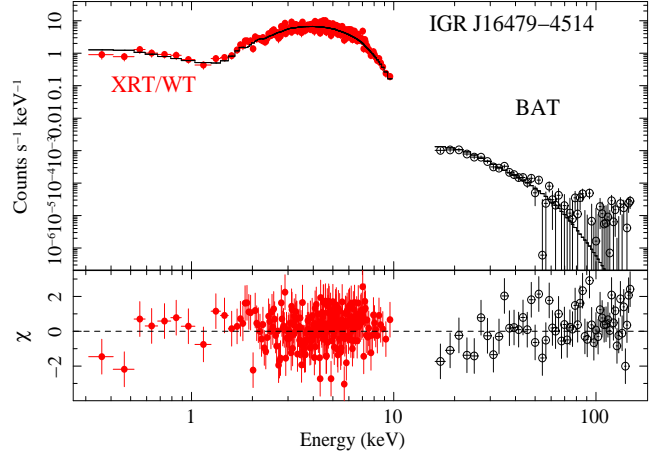}}	
                \caption{Spectroscopy of the 2008 March 19 outburst of
                  IGR~J16479$-$4514. 
		{\bf Top:} data from the second BAT observation  (black
                   empty circles) 
		and simultaneous XRT/WT data (red filled circles) 
		fit with an absorbed power law with a high energy cutoff. 
		{\bf Bottom:} the residuals of the fit in units of standard deviations. 
		   }
                \label{sfxts2:fig:spectrum}
        \end{figure}

The first outburst that triggered the {\it Swift}/BAT since the
beginning of the campaign occurred on 2008 March 19 at 22:44:47 UT,
and then shortly thereafter at 22:59:59 UT
(\cite{Romano2008:sfxts_paperII}),  
when  IGR~J16479 --4514  reached fluxes above
10$^{-9}$~erg~cm$^{-2}$~s$^{-1}$, which
translates into a luminosity of about $6\times10^{37}$ erg s$^{-1}$
(0.5--100\,keV; at a distance of 4.9\,kpc).  
{\it Swift} immediately re-pointed at the target with the narrow-field instruments 
so that, for the first time, an outburst from a SFXT
where a periodicity in the outburst recurrence is unknown
could be observed simultaneously in the 0.2--150\,keV energy band 
(Figure~\ref{sfxts2:fig:lcv_allbands}). 
We observed a highly variable  X--ray emission that spans almost four orders of
magnitude in count rate during the {\it Swift}/XRT observations covering 
a few days before and after the bright peak. 
The XRT spectrum in outburst is hard and highly absorbed (the power-law fit resulted in a 
photon index of 0.98$\pm{0.07}$, and in an absorbing column density of
$\sim$5$\times$10$^{22}$~cm$^{-2}$). 
For the first time a simultaneous broad band X--ray spectrum of
IGR~J16479$-$4514 (also a first in the SFXT class)
was analyzed in the 0.3--100\,keV energy range. 
The source emission is fit well with the spectral models
usually applied to the accreting X--ray pulsars: power laws with a high 
energy cutoff  models (Figure~\ref{sfxts2:fig:spectrum}). 
The resulting parameters are also very similar to those of this
kind of X--ray binary sources.


\section{Disentangling the system geometry in IGR~J11215--5952}

\begin{figure*}[t]
	 	\includegraphics[angle=270,width=15cm]{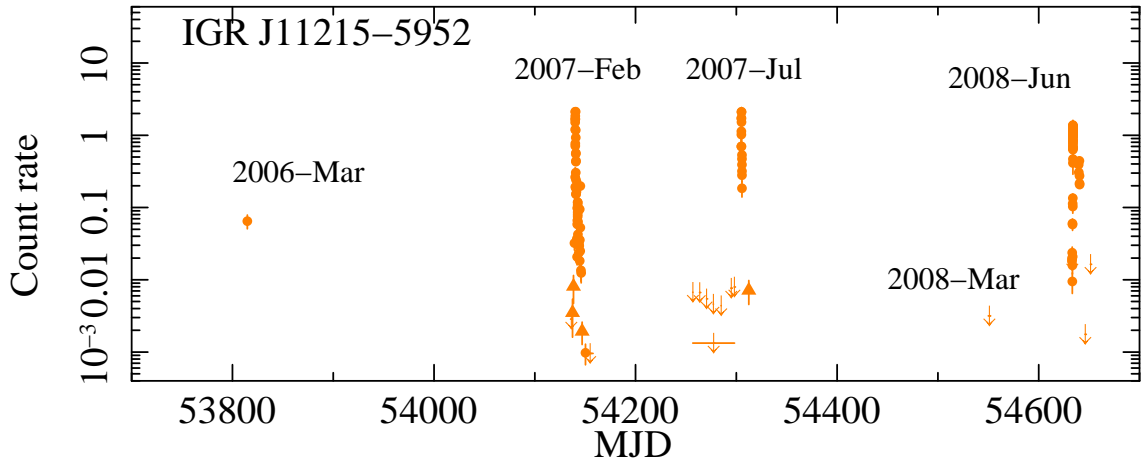}
 		\caption{XRT full light curve of IGR J11215$-$5952
                  (1--10\,keV); the curve is background subtracted,
                  and corrected for pile-up
  (when required), PSF losses, and vignetting.  Downward-pointing
  arrows are 3-$\sigma$ upper limits.
		}
                \label{igr112p3:fig:full_lcv}
	\end{figure*}

Presently, IGR J11215--5952 is the first of two SFXTs displaying
periodic outbursts (the second one being IGR~J18483--0311 which shows
outbursts with a period of 18.52~days, \cite{Sguera2007}). 
While INTEGRAL and RXTE observations have shown
that the outbursts occur with a periodicity of $P=329$ days, thanks to
our {\it Swift} data we have firmly established that the true outburst period is
$\sim 165$ days. In fact, we observed this source with {\it Swift}
several times so that our data sample includes
(Figure~\ref{igr112p3:fig:full_lcv}): 
\begin{enumerate}
\item 2007 February monitoring, intense 23-day long campaign intended
  to study the fifth outburst
  (``periastron'' on February 9, based on a 329\,d period, \cite{Romano2007}), 
 when we could study the evolution of the light
  curve  from outburst onset to almost quiescence, 
  and which constitutes a unique data-set for an
  outburst of a SFXT, thanks to the combination of sensitivity and
  time coverage. 
\item 2007 July (26-day long) monitoring to study the following
  ``apastron''  (July 24, based on $P=329$\,d, \cite{Sidoli2007}) . 
We obtained a ToO for 1.5--2.5\,ks day$^{-1}$ from June 5 to July 31, 
and 17 ks on-source time.
A new, unexpected outburst was observed with
spectral and energetic properties  fully consistent with the ones observed
during the periastron outburst. 
The mean spectrum of the bright peaks can be fit with an absorbed power law model with 
a photon index of $1.00_{-0.14}^{+0.16}$ and an absorbing column of
$(1.04_{-020}^{+0.25}) \times 10^{22}$ cm$^{-2}$. 
This outburst reached luminosities of $\sim 10^{36}$ erg\,s$^{-1}$ (1--10\,keV).
\item 2008 March 25--27 observations to study the $P/4$ (based on
  $P=329$\,d, \cite{Romano2007apastron}). 
No emission was observed down to a 3-$\sigma$ upper limit of $9.1\times10^{-13}$ erg cm$^{-2}$ s$^{-1}$
($\Gamma=1$ and $N_{\rm H}= 10^{22}$ cm$^{-2}$), therefore we
can exclude that the period is 165/2 days.
\item 2008 June monitoring (a 20\,ks GI observation followed by 
 follow-up observations for a total of 12\,ks,
 \cite{Romano2008:11215_2008}) between the predicted outburst
 (``apastron'', based on $P=329$\,d, June  16) and July 4.  
\end{enumerate}

The characteristics of this latest ``apastron'' outburst (2008 June 16) are quite similar to those 
previously observed during the ``periastron'' outburst of 2007
February 9, both in terms of spectral shape and luminosities reached. 
The light curve can 
be modelled with the parameters obtained by \cite{Sidoli2007} for the 
2007 February 9 outburst, although some differences can be observed in its shape. 
The properties of the rise to this new outburst and the comparison with the previous outbursts
allow us to suggest that the true orbital period of IGR J11215$-$5952 
is very likely 164.6~days, and that the orbit 
is eccentric, with the different outbursts produced at the periastron passage,
when the neutron star crosses the inclined equatorial wind from the supergiant companion. 
Based on the observation performed on 2008 March 25--27, we can exclude that 
the period is 165/2 days.

\section{Conclusions}
  
Our almost year-long campaign on this new class of HMXBs has
demonstrated how the unique flexibility of {\it Swift}, 
paired with the high sensitivity of its instruments, 
allowed us to monitor as continuously as possible the onset of the
outbursts in these systems. 
We have obtained multi-wavelength observations for a total of 5
outbursts from 3 sources, which we were able to study in great detail
through wide-band spectroscopy, and time resolved spectroscopy at soft
X-rays.  We have shown (also see \cite{Sidoli2008:igr08})
that common characteristics to this class of objects are
emerging, i.e., outburst lengths well in excess of hours, often with a
multiple peaked structure, dynamic range $\sim3$ orders of magnitude. 
Finally, periodicities are starting to be found.

\acknowledgments
We thank the {\it Swift} team duty scientists and science planners  P.J.\ Brown, M.\ Chester,
E.A.\ Hoversten, S.\ Hunsberger,  C.\ Pagani, J.\ Racusin, and M.C.\ Stroh 
for their dedication and willingness to accomodate our sudden requests
in response to outbursts during this long monitoring effort. 
We also thank the remainder of the {\it Swift} XRT and BAT teams,
J.A.\ Nousek and S.\ Barthelmy in particular, for their invaluable help and support with
the planning and execution of the observing strategy. 
This work was supported in Italy by contracts ASI I/023/05/0 and I/088/06/0, at
PSU by NASA contract NAS5-00136. 
H.A.K. was supported by the {\it Swift } project. 
P.R.\ thanks INAF-IASF Milano and L.S.\ INAF-IASF Palermo, 
for their kind hospitality. 
Italian researchers acknowledge the support of Nature (455, 835-836) and thank
the Editors for increasing the international awareness of the current
critical situation of the Italian Research.


\begin{thebibliography}{99}

\bibitem{Bamba2001} Bamba, A., Yokogawa, J., Ueno, M., et al., 2001, \emph{PASJ}, {\bf 53}, 1179 \\
\bibitem{Bozzo2008} Bozzo, E., Falanga, M., Stella, L., 2008,   \emph{ApJ}, {\bf 683}, 1031 \\
\bibitem{Burrows2005:XRT} Hill, J.~E., Burrows, D.~N., Nousek, J.~A.,   et al., 2005, \emph{SSRv}, {\bf 120}, 165 \\
\bibitem{Cusumano2009} Cusumano, G., Romano, P., Sidoli, L., et al.,  2009, \emph{MNRAS}, in preparation \\ 
\bibitem{Gehrels2004} Gehrels, N., Chincarini, G., Giommi, P., et al., 2004, \emph{ApJ}, {\bf 611}, 1005 \\
\bibitem{Hill04:xrtmodes} Hill, J.~E., Burrows, D.~N., Nousek, J.~A.,   et al., 2004, \emph{SPIE}, {\bf 5165}, 217 \\
\bibitem{zand2005} in't Zand, J.J.M., 2005, \emph{A\&A}, {\bf 441}, L1 \\
\bibitem{Krimm2007:ATel1265} Krimm,  H.~A., Barthelmy, S.~D., Barbier,   L.,   et~al., 2007, \emph{The Astronomer's Telegram}, {\bf 1265}
\bibitem{Negueruela2006} Negueruela, I., Smith, D.M., Reig, P., et al. 2006, in
  \emph{ESA Special Publication}, ed.\ A.\ Wilson, Vol. {\bf 604}, 165 \\
\bibitem{Negueruela2008} Negueruela, I., Torrejon, J.M., Reig, P., et   al., 2008, \emph{AIPC}, {\bf 1010}, 252 \\
\bibitem{Romano2007} Romano, P., Sidoli, L., Mangano, V. et al., 2007, \emph{A\&A},  {\bf 469}, L5  \\
\bibitem{Romano2007apastron} Romano, P., Mangano, V.,  Mereghetti, S., et al., 2007,  \emph{The Astronomer's Telegram}, {\bf 1151} \\
\bibitem{Romano2008:sfxts_paperII} Romano, P., Sidoli, L., Mangano, V., et al., 2008, \emph{ApJ},  {\bf 680}, L137   [Paper~II]    \\
\bibitem{Romano2008:atel1659} Romano, P., Guidorzi, C., Sidoli, L., et al., 2008, \emph{The Astronomer's Telegram}, {\bf 1659} \\
\bibitem{Romano2008:atel1697} Romano, P., Cusumano, G., Sidoli, L., et al., 2008, \emph{The Astronomer's Telegram}, {\bf 1697} \\
\bibitem{Romano2008:11215_2008} Romano, P., Sidoli, L.,  Cusumano, G., et al., 2009,  \emph{ApJ}, in press, [arXiv:0902.1985]  \\
\bibitem{Sguera2007} Sguera, V., Hill, A. B., Bird, A. J., et al., 2007, \emph{A\&A}, {\bf  467}, 249 \\
\bibitem{Sidoli2007} Sidoli, L., Romano, P., Mereghetti, S., et al.,   2007, \emph{A\&A}, {\bf 476}, 1307 \\
\bibitem{Sidoli2008:sfxts_paperI} Sidoli, L., Romano, P., Mangano, V., et al., 2008, \emph{ApJ}, {\bf 687}, 1230      [Paper~I] \\
\bibitem{Sidoli2008:sfxts_paperIII} Sidoli, L., Romano, P., Mangano, V., et al., 2009, \emph{ApJ}, {\bf 690}, 120    [Paper~III] \\
\bibitem{Sidoli2008:cospar} Sidoli, L., 2008, Proc. of the COSPAR Assembly 2008, {\bf 37}, 2892, [arXiv:0809.3157] \\
\bibitem{Sidoli2008:igr08} Sidoli, L., Romano, P., Cusumano, G., et
  al.,  2009, in proceedings of \emph{7th INTEGRAL Workshop}, \pos{PoS(Integral08)084},  [arXiv:0810.5446] \\
\bibitem{Smith1998:17391-3021} Smith, D.~M.,  Main, D., Marshall, F., et al., 1998, \emph{ApJL}, {\bf 501}, 181   \\
\bibitem{Swank2007} Swank, J.H., Smith, D.M., Markwardt, C.B., 2007, \emph{The Astronomer's Telegram}, {\bf 999}\\
\bibitem{Walter2007} Walter, R., \& Zurita Heras, J., 2007, \emph{A\&A}, {\bf 476}, 335 \\


\end{thebibliography}
\end{document}